\documentclass[11pt]{article}

\usepackage[margin=1in]{geometry}
\usepackage{amsmath,amssymb,amsfonts,amsthm}
\usepackage{booktabs}
\usepackage{hyperref}
\usepackage[numbers]{natbib}
\newcommand{\R}{\mathbb{R}}
\newcommand{\E}{\mathbb{E}}

\newtheorem{theorem}{Theorem}[section]
\newtheorem{proposition}[theorem]{Proposition}
\newtheorem{corollary}[theorem]{Corollary}
\theoremstyle{definition}
\newtheorem{remark}[theorem]{Remark}
\DeclareMathOperator{\df}{df}

\title{Spectral Regularization Beyond Ridge Regression}

\author{
Kathryn N. Vasilaky\thanks{Department of Economics and Noyce School of Applied Computing, California Polytechnic State University, San Luis Obispo, CA (\texttt{kvasilak@calpoly.edu}). Additional affiliations: Columbia Climate School, Columbia University; Center for Effective Global Action (CEGA), UC Berkeley; Abdul Latif Jameel Poverty Action Lab (J-PAL), MIT; and Innovation Growth Lab.}
\and
Walter Vasilaky\thanks{Independent Researcher.(\texttt{wvasilaky@gmail.com})}
}

\date{}

\begin{document}

\maketitle

\begin{abstract}
We study Inclusive Ridge (IR), a two-parameter spectral regularization family that contains ridge regression as well as iterated Tikhonov regularization.  We derive its exact finite-sample coefficient risk and establish necessary and sufficient conditions for increasing or decreasing the additional filter-shape parameter around the ridge baseline to locally lower risk. We show that oracle-tuned ridge is generically suboptimal within the IR family. Under joint divergence of the penalty and filter order, we show that a nondegenerate spectral limit exists precisely when their ratio converges to a positive finite constant, yielding a continuous exponential filter. Along the fixed effective-scale path, we also obtain an exact risk-difference comparison and explicit sufficient conditions under which both higher-order estimators and the exponential limiting filter strictly outperform oracle-tuned ridge. In this regime, IR becomes progressively less sensitive to perturbations in the absolute penalty parameter while retaining sensitivity to its effective spectral scale, and a plug-in coefficient-risk first-order condition coincides with a degrees-of-freedom-adjusted discrepancy equation. Numerical examples based on the Fredholm inverse problem, tomographic reconstruction, a regression design, and  synthetic difference-in-differences donor-weight estimation illustrate both the potential gains and the limitations of feasible tuning, distinguishing oracle improvements from gains attainable with estimated tuning parameters. 
\end{abstract}


% --- PASTE \section{Introduction} RIGHT BELOW THIS LINE ---

\section{Introduction}

Ridge regression \cite{Hoerl1970} stabilizes ill-conditioned linear estimation by applying the rational spectral filter $\sigma_j^2 / (\sigma_j^2 + \zeta)$ to singular direction $\sigma_j$. A single penalty $\zeta$ determines shrinkage over the entire spectrum. When signal energy varies substantially across directions, this restriction can be costly for coefficient recovery.

We examine the two-parameter filter
\begin{equation}\label{eq:ir_filter}
f_{\zeta,\eta}(s) = 1 - \left(\frac{\zeta}{\zeta + s}\right)^\eta, \qquad \zeta > 0, \quad \eta > 0, \quad s > 0,
\end{equation}
evaluated at squared singular values \(s = \sigma_j^2\). Ridge corresponds to $\eta = 1$. Positive integer orders $\eta$ recover stationary iterated Tikhonov regularization \cite{Engl1996, Hanke1998},  while arbitrary positive orders define a continuous spectral extension. Rather than introducing iterated Tikhonov or exponential regularization, our contribution lies in establishing risk improvement over oracle-tuned ridge, characterizing filter behavior on both sides of $\eta = 1$, proving joint-scaling convergence to the exponential filter, and linking coefficient-risk tuning to a discrepancy equation.

Our principal finite-sample result derives an explicit first-order spectral contrast at an interior ridge optimum, providing necessary and sufficient conditions for increasing or decreasing $\eta$ around $\eta = 1$ to strictly lower coefficient risk. Because this contrast vanishes only on a set of Lebesgue measure zero, oracle-tuned ridge is generically suboptimal within the broader family. However, because this is an oracle guarantee, estimating a second hyperparameter from noisy data introduces tuning variance that can erode these gains when the spectral contrast is weak.

The family also connects discrete iteration with continuous spectral regularization. Along the linear path $\zeta = \kappa \eta$ with $\kappa \in (0, \infty)$, the filter converges as $\eta \to \infty$ to the exponential filter $1 - \exp(-s/\kappa)$, so the effective scale is $\kappa$ rather than the absolute penalty $\zeta$. Beyond local perturbations of $\eta = 1$, we derive an exact risk-difference decomposition along the fixed effective-scale path $\zeta = \eta\zeta^*$ and obtain explicit sufficient conditions under which both higher-order estimators ($\eta > 1$) and the exponential limiting filter at $\kappa = \zeta^*$ strictly outperform oracle-tuned ridge. We also analyze the scale distinction between $\kappa$ and $\zeta$ and obtain a degrees-of-freedom-adjusted residual identity for plug-in coefficient-risk tuning of the limiting filter.

The numerical section prioritizes coefficient recovery in the Phillips and tomographic reconstruction inverse problems \cite{Hansen2010}, while treating regression and synthetic difference-in-differences \cite{Arkhangelsky2021} as tests of the method's boundaries. While IR’s additional spectral flexibility does not confer a universal advantage and can default to ridge, these experiments demonstrate precisely when it yields meaningful gains in coefficient estimation.
\section{Estimator and finite-sample risk}
\subsection{Model and spectral definition}
Let
\begin{equation}\label{eq:model}
 y=X\beta+\varepsilon,\qquad
 \varepsilon\sim N(0,\sigma_\varepsilon^2 I_n),
\end{equation}
where $X\in\R^{n\times p}$ has full column rank, $n\ge p$, and
$X=U\operatorname{diag}(\sigma_1,\ldots,\sigma_p)V^\top$ with
$\sigma_1\ge\cdots\ge\sigma_p>0$. The initial target $\beta_0\in\R^p$
is fixed independently of $\varepsilon$. Define
$a_j=v_j^\top(\beta-\beta_0)$, $z_j=u_j^\top(y-X\beta_0)$, and
$r_j(\zeta)=\zeta/(\zeta+\sigma_j^2)$. The IR estimator is
\begin{equation}\label{eq:estimator}
 \widehat\beta_{\zeta,\eta}
 =\beta_0+\sum_{j=1}^p
 [1-r_j(\zeta)^\eta]\frac{z_j}{\sigma_j}v_j.
\end{equation}
For integer $\eta=k\ge1$, this equals the $k$th iterate of
\begin{equation}\label{eq:recursion}
 \beta^{(k)}
 =(X^\top X+\zeta I)^{-1}(X^\top y+\zeta\beta^{(k-1)}),
 \qquad\beta^{(0)}=\beta_0.
\end{equation}
For noninteger $\eta$, equation \eqref{eq:estimator}, rather than a
fractional iteration of \eqref{eq:recursion}, is the definition.

\begin{proposition}[Exact coefficient risk]\label{prop:risk}
Under \eqref{eq:model}, with fixed $\beta_0$,
\begin{equation}\label{eq:risk}
 R(\zeta,\eta):=\E\|\widehat\beta_{\zeta,\eta}-\beta\|^2
 =\sum_{j=1}^p\left[
 \frac{\sigma_\varepsilon^2}{\sigma_j^2}(1-r_j^\eta)^2
 +a_j^2r_j^{2\eta}\right].
\end{equation}
\end{proposition}
\begin{proof}
Since $z_j=\sigma_ja_j+u_j^\top\varepsilon$,
the error in direction $v_j$ equals
$-r_j^\eta a_j+(1-r_j^\eta)u_j^\top\varepsilon/\sigma_j$.
Orthogonality and zero-mean noise yield \eqref{eq:risk}.
\end{proof}
\begin{remark}
A fixed target $\beta_0$ changes squared bias without changing the
variance term. If $\beta_0$ is estimated from the same observations,
its covariance with the noise must be included and
\eqref{eq:risk} need not apply unchanged.
Also, $0<f_{\zeta,\eta}(\sigma_j^2)<1$ bounds amplification in each
direction relative to OLS; it does not by itself imply an improved
matrix condition number.
\end{remark}
\begin{remark}[Computational complexity]\label{rem:complexity}
Evaluating \eqref{eq:estimator} for arbitrary real $\eta > 0$ relies on a thin singular value decomposition of $X$, which costs $O(\min(np^2, n^2p))$ operations. Once the singular components are computed, however, evaluating the estimator, its exact risk \eqref{eq:risk}, or the limiting filter \eqref{eq:exp} across any grid of $(\zeta, \eta)$ requires only $O(p)$ operations. For positive integer orders $\eta = k$, the proximal recursion \eqref{eq:recursion} avoids the SVD altogether via $k$ linear solves, while large-scale non-integer implementations can substitute randomized SVD or Krylov subspace approximations.
\end{remark}

\subsection{When does the second parameter improve upon ridge?}
Write $\zeta^*\in(0,\infty)$ for an interior global minimizer of
$R(\zeta,1)$ and set
\begin{equation}
 W_j=\frac{\zeta^*\log(1+\sigma_j^2/\zeta^*)}
 {(\zeta^*+\sigma_j^2)^2},\qquad
 P_j=\frac{W_j}{\sum_{\ell=1}^pW_\ell}.
\end{equation}
\begin{proposition}[Direction of local improvement]\label{prop:direction}
At an interior ridge optimum,
\begin{equation}\label{eq:derivative}
 \left.\frac{\partial R(\zeta^*,\eta)}{\partial\eta}\right|_{\eta=1}
 =2\sum_{j=1}^pW_j
 (\sigma_\varepsilon^2-\zeta^*a_j^2).
\end{equation}
Consequently, increasing $\eta$ from one strictly decreases risk
locally if and only if
\begin{equation}\label{eq:criterion}
 \sigma_\varepsilon^2<\zeta^*\sum_jP_ja_j^2.
\end{equation}
If the strict inequality is reversed, decreasing $\eta$ from one
strictly decreases risk locally.
\end{proposition}
\begin{proof}
Differentiate \eqref{eq:risk} using
$\partial_\eta r_j^\eta=r_j^\eta\log r_j$. At $\eta=1$,
\[
 \partial_\eta R
 =2\sum_jr_j(-\log r_j)
 \left[\frac{\sigma_\varepsilon^2(1-r_j)}{\sigma_j^2}
 -_j^2r_j\right].
\]
Substitution of $r_j=\zeta^*/(\zeta^*+\sigma_j^2)$ gives
\eqref{eq:derivative}. The sign conditions follow because every $W_j>0$.
\end{proof}
\begin{corollary}[Conditional strict dominance]\label{cor:dominance}
If the two sides of \eqref{eq:criterion} are unequal, then
\[
 \inf_{\zeta>0,\eta>0}R(\zeta,\eta)
 <\min_{\zeta>0}R(\zeta,1).
\]
The improvement is achieved in a neighborhood of
$(\zeta^*,1)$ by changing $\eta$ alone.
\end{corollary}
\begin{proof}
The derivative in Proposition \ref{prop:direction} is nonzero.
Move a sufficiently small positive or negative distance from $\eta=1$
in its decreasing direction.
\end{proof}

\begin{proposition}[Monotone spectral signal]\label{prop:monotone}
Assume at least two distinct singular values and an interior ridge
optimum. If $a_j^2$ is nondecreasing as a function of $\sigma_j^2$
and is not constant, then \eqref{eq:criterion} holds.
If $a_j^2$ is nonincreasing and is not constant, the reverse
strict inequality holds.
\end{proposition}
\begin{proof}
The ridge first-order condition is
$\sum_jq_j(\zeta^*a_j^2-\sigma_\varepsilon^2)=0$,
where $q_j=\sigma_j^2/(\zeta^*+\sigma_j^2)^3>0$.
Set $t_j=\sigma_j^2/\zeta^*$ and
$g(t)=(1+t)\log(1+t)/t$. Then $W_j=\zeta^*q_jg(t_j)$
and $g$ is strictly increasing. The derivative
\eqref{eq:derivative} is therefore minus a positive multiple of
the weighted covariance between $g(t_j)$ and $a_j^2$.
The weighted rearrangement identity proves the sign claim
when both sequences are similarly or oppositely ordered.
\end{proof}

\begin{proposition}[Geometric characterization of the exceptional set]
\label{prop:exceptional_set}
Let $X\in\mathbb{R}^{n\times p}$ have full column rank,
where $p\geq 2$, and let
$s_j=\sigma_j^2>0$, $j=1,\ldots,p$, denote the squared
singular values of $X$. Assume that at least two of the
$s_j$ are distinct and that the noise variance
$v=\sigma_\varepsilon^2>0$ is fixed.

Setting $\beta_0=0$ without loss of generality, for a signal $\beta\in\mathbb{R}^p$, let
$\alpha_j=v_j^\top\beta$, and define the spectral risk
\[
R(\zeta,\eta;\beta)
=
\sum_{j=1}^p
\left[
\frac{v}{s_j}
\left\{1-
\left(\frac{\zeta}{\zeta+s_j}\right)^\eta
\right\}^2
+
\alpha_j^2
\left(\frac{\zeta}{\zeta+s_j}\right)^{2\eta}
\right].
\]

Define the exceptional set
\[
\mathcal E
=
\left\{
\beta\in\mathbb{R}^p:
\begin{array}{l}
\text{there exists }\zeta>0\text{ such that}\\
\partial_\zeta R(\zeta,1;\beta)=0,\\
\partial_\eta R(\zeta,1;\beta)=0
\end{array}
\right\}.
\]
Then $\mathcal E$ has $p$-dimensional Lebesgue
measure zero.

In particular, for Lebesgue-almost every signal
$\beta$, every interior minimizer $\zeta^*$ of
$R(\zeta,1;\beta)$ satisfies
\[
\left.
\frac{\partial R(\zeta^*,\eta;\beta)}
{\partial\eta}
\right|_{\eta=1}\neq 0.
\]
\end{proposition}
\begin{proof}
    See Appendix.
\end{proof}

\begin{remark}[Isotropic signals and the exceptional boundary]
\label{rem:isotropic_boundary}
If $\alpha_j^2=a>0$ for all $j$, ridge with
$\zeta^*=\sigma_\varepsilon^2/a$ minimizes
each spectral risk contribution separately
among all filters taking values in $[0,1]$.
Thus isotropy is sufficient for global
ridge optimality.

More generally, equality in the first-order
condition \eqref{eq:criterion} need not imply isotropy or
global optimality. Nevertheless, under the
nondegeneracy assumptions of
Proposition~\ref{prop:exceptional_set},
the set of signals for which this equality
holds at an interior ridge stationary point
has Lebesgue measure zero.

Consequently, for almost every signal,
the risk derivative with respect to $\eta$
at an interior ridge optimum is nonzero.
The sign of this derivative determines
whether increasing $\eta$ locally improves
or worsens the risk.
\end{remark}
\subsection{Comparison along a fixed effective-scale path}
For $\eta>1$, define
$t_j=\sigma_j^2/\zeta^*$,
$r_j=(1+t_j)^{-1}$ and
$b_j=(1+t_j/\eta)^{-\eta}$.
Since $0<b_j<r_j$,
\begin{align}
 R(\eta\zeta^*,\eta)-R(\zeta^*,1)
 &=\frac{\sigma_\varepsilon^2}{\zeta^*}
 \sum_j\frac{(r_j-b_j)(2-r_j-b_j)}{t_j}\notag\\
 &\quad-\sum_j a_j^2(r_j-b_j)(r_j+b_j).
 \label{eq:pathdiff}
\end{align}
This exact identity provides a finite-order comparison with
optimally tuned ridge. Its two terms show the variance cost and
squared-bias benefit of spectral release. The analogous exponential
comparison follows by replacing $b_j$ with $e^{-t_j}$.
Neither comparison asserts improvement for every signal.

\section{Critical scaling and exponential regularization}
\begin{proposition}[Critical scaling]\label{prop:limit}
Suppose $\eta\to\infty$, $\zeta_\eta\to\infty$ and
$\zeta_\eta/\eta\to\kappa\in[0,\infty]$.
For each fixed $s>0$,
\[
 \left(1+\frac{s}{\zeta_\eta}\right)^{-\eta}
 \longrightarrow
 \begin{cases}
 0,&\kappa=0,\\
 e^{-s/\kappa},&0<\kappa<\infty,\\
 1,&\kappa=\infty.
 \end{cases}
\]
If a full sequence of these survival factors converges to a
value strictly between zero and one at any fixed $s>0$,
then $\zeta_\eta/\eta$ necessarily converges to a finite,
strictly positive constant.
\end{proposition}
\begin{proof}
Write the logarithm of the survival factor as
$-\eta\log(1+s/\zeta_\eta)$.
Because $\zeta_\eta\to\infty$,
$\log(1+s/\zeta_\eta)/(s/\zeta_\eta)\to1$.
The stated cases follow from the limit of $\eta/\zeta_\eta$.
Conversely, if the survival factor converges to $u\in(0,1)$,
then $-\eta\log(1+s/\zeta_\eta)\to\log u$,
so $\eta/\zeta_\eta\to-\log u/s\in(0,\infty)$.
\end{proof}
With the convention $\zeta_\eta/\eta\to\kappa>0$,
the limiting estimator is
\begin{equation}\label{eq:exp}
 \widehat\beta_\kappa^\infty
 =\beta_0+\sum_{j=1}^p
 (1-e^{-\sigma_j^2/\kappa})\frac{z_j}{\sigma_j}v_j.
\end{equation}
Its coefficient risk is
\begin{equation}\label{eq:exprisk}
 R_\infty(\kappa)=\sum_{j=1}^p
 \left[
 a_j^2e^{-2\sigma_j^2/\kappa}
 +\frac{\sigma_\varepsilon^2}{\sigma_j^2}
 (1-e^{-\sigma_j^2/\kappa})^2
 \right].
\end{equation}
The $q$-power scaling $\zeta=\kappa\eta^q$ yields OLS
for $q<1$, \eqref{eq:exp} for $q=1$, and $\beta_0$ for $q>1$.
These are fixed-design, fixed-$p$ limits; growing-dimensional
claims require additional uniform assumptions.

The limit is the solution at time $t=1/\kappa$ of the linear
gradient flow
$\dot\beta(t)=X^\top(y-X\beta(t))$ initialized at $\beta_0$ \cite{Showalter1974, Buhlmann2003}.
Finite-order IR is instead the discrete proximal iteration
\eqref{eq:recursion} for integer orders.
\begin{remark}[Connection to $L_2$-boosting]\label{rem:l2_boosting}
For positive integer orders $\eta = m \in \mathbb{N}$ and $\beta_0 = 0$, the IR filter \eqref{eq:ir_filter} coincides with $m$-step $L_2$-boosting using a linear ridge base learner~\cite{Buhlmann2003}. Viewing this family through the continuous spectral parameterization \eqref{eq:estimator} offers three advantages over standard iterative boosting implementations:
\begin{enumerate}
    \item \emph{Continuous filter orders:} Allowing arbitrary $\eta > 0$ creates a smooth risk surface for joint optimization and enables softened filters ($0 < \eta < 1$) that are unattainable with discrete boosting steps.
    \item \emph{Single-SVD amortization:} Rather than performing sequential matrix solves across $m$ boosting iterations for each candidate penalty, a single thin SVD reduces evaluation across any grid of $(\zeta, \eta)$---as well as direct plug-in tuning of the exponential limit \eqref{eq:exp}---to $O(p)$ scalar operations.
    \item \emph{Asymptotic unification:} Proposition~\ref{prop:limit} establishes that discrete ridge boosting converges to continuous-time gradient flow if and only if the penalty $\zeta$ and iteration count $m$ grow at the same rate.
\end{enumerate}
\end{remark}

\begin{corollary}[Distribution of the limiting estimator]\label{cor:limit_dist}
Under \eqref{eq:model} with fixed $\beta_0$ and $\kappa > 0$, the limiting exponential estimator \eqref{eq:exp} has the exact Gaussian distribution
\begin{equation}\label{eq:limit_dist}
 \widehat\beta_\kappa^\infty \sim N(\mu_\kappa, \Sigma_\kappa),
\end{equation}
where
\begin{align}
 \mu_\kappa &= \beta_0 + \sum_{j=1}^p \left(1 - e^{-\sigma_j^2/\kappa}\right) a_j v_j, \label{eq:limit_mean}\\
 \Sigma_\kappa &= \sigma_\varepsilon^2 \sum_{j=1}^p \left(\frac{1 - e^{-\sigma_j^2/\kappa}}{\sigma_j}\right)^2 v_j v_j^\top. \label{eq:limit_cov}
\end{align}
\end{corollary}
\begin{proof}
Since $z_j = u_j^\top(y - X\beta_0) = \sigma_j a_j + u_j^\top \varepsilon$ with $\varepsilon \sim N(0, \sigma_\varepsilon^2 I_n)$, the coordinates $z_1, \ldots, z_p$ are independent with $z_j \sim N(\sigma_j a_j, \sigma_\varepsilon^2)$. Because $\widehat\beta_\kappa^\infty$ in \eqref{eq:exp} is an affine transformation of $(z_1, \ldots, z_p)$, it is Gaussian with mean \eqref{eq:limit_mean} and covariance \eqref{eq:limit_cov}, using orthonormality of $v_1, \ldots, v_p$.
\end{proof}

\begin{remark}[Sensitivity]
For $s>0$,
\[
 \frac{\partial f_{\zeta,\eta}(s)}{\partial\zeta}
 =-\frac{\eta s}{(\zeta+s)^2}
 \left(\frac{\zeta}{\zeta+s}\right)^{\eta-1}.
\]
At $\zeta=\kappa\eta$ this is $O(\eta^{-1})$ for
fixed $s,\kappa>0$. Thus a fixed \emph{absolute} change
in $\zeta$ has a diminishing effect. A proportional change
in $\zeta$ changes the effective scale $\kappa$ and need
not have a diminishing effect. This is a parameterization
sensitivity result, not a general assertion that the
optimally tuned risk surface becomes flat.
\end{remark}
\section{Feasible tuning}
\subsection{Plug-in coefficient risk and discrepancy identity}
The following uses the time parameter $t=1/\kappa$ to avoid
confusing the exponential rate with the penalty scale.
Let $u_j(t)=e^{-t\sigma_j^2}$ and
$\widehat\sigma_\varepsilon^2>0$. Define the unbiased
per-direction estimate of squared signal energy
$\widehat a_j^2=(z_j^2-\widehat\sigma_\varepsilon^2)/\sigma_j^2$,
and
\begin{equation}\label{eq:plugin}
 \widehat R_\infty(t)=
 \sum_j\left[u_j(t)^2\widehat a_j^2+
 \frac{\widehat\sigma_\varepsilon^2}{\sigma_j^2}
 (1-u_j(t))^2\right].
\end{equation}
The estimated squared coefficients may be negative;
unbiased plug-in risk and nonnegative truncated
coefficient estimates are different procedures.

\begin{proposition}[Risk stationarity and discrepancy]\label{prop:discrepancy}
Let $\df(t)=\sum_j(1-u_j(t))$. Then
\begin{equation}\label{eq:derivplugin}
 \frac{d\widehat R_\infty(t)}{dt}
 =-2\left[\sum_j u_j(t)^2z_j^2-
 \widehat\sigma_\varepsilon^2(p-\df(t))\right].
\end{equation}
Thus every interior stationary point satisfies
\begin{equation}\label{eq:stationary}
 \sum_j u_j(t)^2z_j^2
 =\widehat\sigma_\varepsilon^2(p-\df(t)).
\end{equation}
If $n>p$ and
$\widehat\sigma_\varepsilon^2=
 \|(I-UU^\top)y\|^2/(n-p)$,
this is equivalent to
\begin{equation}\label{eq:discrep}
 \|y-X\widehat\beta^\infty_{1/t}\|^2
 =\widehat\sigma_\varepsilon^2(n-\df(t)).
\end{equation}
\end{proposition}
\begin{proof}
Differentiate \eqref{eq:plugin}, using
$u_j'=-\sigma_j^2u_j$, and cancel the two
$\widehat\sigma_\varepsilon^2u_j^2$ terms.
This yields \eqref{eq:derivplugin}.
The residual has in-range coordinates $u_j(t)z_j$
and an unchanged orthogonal component
$(I-UU^\top)y$, giving \eqref{eq:discrep}.
\end{proof}
\begin{remark}
Stationarity is not sufficient for global risk minimization;
the objective and its boundary values must also be compared.
The equivalence is an identity for the specified plug-in
\emph{coefficient} risk and residual-space noise estimate,
not a general identity for arbitrary regularizers.
\end{remark}

\subsection{Generalized cross-validation}
For fixed $\beta_0$, let
$h_j=1-r_j^\eta$ and
$H_{\zeta,\eta}=U\operatorname{diag}(h_j)U^\top$.
The fit is affine in $y$ if $\beta_0\ne0$,
with derivative $H_{\zeta,\eta}$.
Its residual is
$y-X\widehat\beta_{\zeta,\eta}
 =(I-UU^\top)y+\sum_jr_j^\eta z_ju_j$.
Consequently, a GCV-style criterion \cite{Golub1979}  is
\begin{equation}\label{eq:gcv}
 \operatorname{EGCV}(\zeta,\eta)=
 \frac{n\left[\|(I-UU^\top)y\|^2+
 \sum_{j=1}^p r_j^{2\eta}z_j^2\right]}
 {\left(n-p+\sum_{j=1}^p r_j^\eta\right)^2}.
\end{equation}
The criterion targets predictive fit rather than coefficient risk.
No claim of uniform superiority from tuning two parameters is
made without an accompanying selection-error analysis.

\subsection{Picard-tail selection and its limitations}
\label{prop:picard}
When an inverse problem has no orthogonal residual space, noise estimation requires additional information or a spectral signal-tail assumption~\cite{Hansen1990, Hansen2010, OLeary2001, Levin-Meltzer2017}. To operationalize the limiting exponential estimator $\widehat{\beta}_{\kappa}^{\infty}$ in \eqref{eq:exp} without cross-validation, we decouple the signal and noise subspaces using the data-driven tail-energy rule of Levin and Meltzer~\cite{Levin-Meltzer2017}. Setting $\beta_0 = 0$ and writing $c_j = u_j^\top y$ for $j = 1, \ldots, r$ with $r = \operatorname{rank}(X)$, with $r = p$ under our full-rank assumption, define the averaged tail energy
\begin{equation}\label{eq:tail_energy}
\mathcal{V}(k) = \frac{1}{r - k + 1} \sum_{j=k}^{r} c_j^2.
\end{equation}
Given a window $h \ge 1$ and relative tolerance $\delta > 0$, the empirical Picard index $\widehat{\iota}$ is selected by the stopping rule
\begin{equation}\label{eq:picard_index}
\widehat{\iota} = \min\left\{ k \ge 1 : \frac{|\mathcal{V}(k+h) - \mathcal{V}(k)|}{\mathcal{V}(k)} < \delta \right\}.
\end{equation}
The noise variance is then estimated at the boundary by $\widehat{\sigma}_{\varepsilon}^2 = \mathcal{V}(\widehat{\iota})$, while the squared signal coefficients in \eqref{eq:exprisk} are replaced by the truncated plug-in proxies
\begin{equation}\label{eq:picard_signal}
\widehat{a}_j^2 =
\begin{cases}
c_j^2 / \sigma_j^2, & j < \widehat{\iota}, \\
0, & j \ge \widehat{\iota}.
\end{cases}
\end{equation}
Minimizing the resulting plug-in coefficient risk $\sum_{j=1}^r \big[\widehat{a}_j^2 e^{-2\sigma_j^2/\kappa} + (\widehat{\sigma}_{\varepsilon}^2/\sigma_j^2)(1 - e^{-\sigma_j^2/\kappa})^2\big]$ over $\kappa > 0$ yields a data-driven scale estimate $\widehat{\kappa}$. (Notice that unlike the unbiased estimate $\widehat{a}_j^2 = (z_j^2 - \widehat{\sigma}_\varepsilon^2)/\sigma_j^2$ in Section~4.1, \eqref{eq:picard_signal} omits the noise-bias subtraction for $j < \widehat{\iota}$ and sets all coordinates $j \ge \widehat{\iota}$ to zero.)

This procedure is appropriate only to the extent that the tail assumption holds. In regression with $n > r$, the residual-space estimator is available under \eqref{eq:model} and can be used to test or calibrate the spectral-tail assumption. Accordingly, feasible plug-in results should be reported separately from oracle tuning.
\section{Applications and Experiments}\label{sec:applications}

We examine Inclusive Ridge (IR) in four settings with different sources of ill-conditioning and different evaluation criteria: (1) the Phillips inverse problem, which isolates the spectral mechanism under oracle coefficient risk; (2) tomographic reconstruction, which tests how filter-shape gains depend on signal regularity; (3) the Ames house-price regression, which contrasts Picard-tail and residual-space noise estimation under feasible tuning; and (4) synthetic difference-in-differences (SDID), which tests whether regularizing first-stage donor weights improves downstream treatment-effect estimation. Throughout, oracle comparisons quantify the attainable gain of a filter family and are kept distinct from data-driven selection rules. Python scripts of the numerical examples are provided https://github.com/vasilaky/inclusive-ridge.

\subsection{The Phillips Fredholm inverse problem}
The Phillips problem~\cite{Hansen2010} is a standard discretization of a
first-kind Fredholm integral equation whose rapidly decaying singular values
make direct inversion unstable. Using the $n=400$ discretization, we write
$A=U\Sigma V^{\top}$, $\alpha_{j}=v_{j}^{\top}\boldsymbol{\beta}$, and
$\mathbf{y}=A\boldsymbol{\beta}+\boldsymbol{\varepsilon}$ with
$\boldsymbol{\varepsilon}\sim\mathcal{N}(0,\sigma_{\varepsilon}^{2}I)$, where
$\sigma_{\varepsilon}=0.05\,\|A\boldsymbol{\beta}\|/\sqrt{n}$ corresponds to a
noise level of 5\% of the noiseless response norm. With $s_{j}=\sigma_{j}^{2}$,
the ridge, finite-IR, and limiting exponential filters are
\begin{equation}
f_{j}^{R}=\frac{s_{j}}{s_{j}+\zeta},\qquad
f_{j}^{IR}=1-\left(\frac{\zeta}{\zeta+s_{j}}\right)^{\eta},\qquad
f_{j}^{\infty}=1-e^{-s_{j}/\kappa},
\label{eq:phillips_filters}
\end{equation}
with reconstruction risk
\begin{equation}
R(f)=\sum_{j=1}^{p}\left[\frac{\sigma_{\varepsilon}^{2}}{s_{j}}f_{j}^{2}
+\alpha_{j}^{2}(1-f_{j})^{2}\right].
\label{eq:phillips_risk}
\end{equation}
Tuning parameters minimize the expected oracle risk~\eqref{eq:phillips_risk},
which isolates filter shape from empirical parameter selection and does not
depend on any particular noise realization.

\begin{table}[htbp]
\centering
\caption{Phillips problem ($n=400$): oracle coefficient MSE,
$R(f)/n$ from~\eqref{eq:phillips_risk}. Percentage gains use optimally tuned
ridge as the reference. The finite-IR search is restricted to $\eta\le 200$.}
\label{tab:phillips}
{\small
\begin{tabular}{lcccc}
\toprule
Estimator & $\zeta^{*}$ & $\eta^{*}$ or $\kappa^{*}$ & MSE & Gain (\%)\\
\midrule
Ridge       & 0.15585681 & $\eta=1$             & $3.94794144\times10^{-5}$ & ---\\
Finite IR   & 61.047233  & $\eta=200$           & $2.69968249\times10^{-5}$ & 31.618\\
Limiting IR & ---        & $\kappa=0.30637617$  & $2.69561188\times10^{-5}$ & 31.721\\
\bottomrule
\end{tabular}}
\end{table}

Finite IR lowers oracle MSE by approximately 31.6\% relative to optimally
tuned ridge, and the limiting exponential filter attains a slightly lower risk
still (31.7\%). The finite-IR risk keeps decreasing as $\eta$ grows, so its
optimum lies at the search bound $\eta=200$. In this problem, the
exponential limit is the best member of the family along the scaling path.
Consistent with this, the finite-IR optimum satisfies
$\zeta^{*}/\eta^{*}=0.30523617$, within 0.4\% of the independently optimized
limiting scale $\kappa^{*}=0.30637617$. This agreement illustrates the
critical scaling $\zeta=\kappa\eta$. It also shows that $(\zeta,\eta)$ are
only weakly identified individually when both are large: the risk is
essentially flat along the ray $\zeta/\eta=\kappa^{*}$.

\subsection{Tomographic reconstruction: signal regularity}\label{sec:tomography}

We next evaluate a parallel-beam Radon transform~\cite{Hansen2010} applied to a $32\times32$ image across 24 projection angles, yielding $1{,}104$ measurements, $1{,}024$ unknowns, and an estimated condition number of approximately $7\times10^8$. We compare two targets with contrasting spectral decay: a smooth two-bump image and the Shepp--Logan phantom~\cite{SheppLogan1974}, whose sharp boundaries induce slower spectral decay. For relative Gaussian noise levels of $1\%$ and $5\%$, we generate 20 realizations and compare oracle-tuned ridge, finite IR ($\eta\in\{2,5\}$), and the limiting exponential estimator using relative reconstruction error within the row space of the forward operator.

\begin{table}[htbp]
\centering
\caption{Tomographic reconstruction: relative error under oracle tuning, averaged over 20 noise realizations.}
\label{tab:tomography}
\small
\begin{tabular}{lccccc}
\toprule
Signal & Noise & Ridge & IR$_2$ & IR$_5$ & IR$_\infty$\\
\midrule
Smooth & $1\%$ & 0.040 & 0.033 & 0.029 & 0.028\\
Smooth & $5\%$ & 0.113 & 0.090 & 0.077 & 0.068\\
Shepp--Logan & $1\%$ & 0.114 & 0.112 & 0.112 & 0.113\\
Shepp--Logan & $5\%$ & 0.204 & 0.196 & 0.198 & 0.198\\
\bottomrule
\end{tabular}
\end{table}

The results show a strong dependence on signal regularity. For the smooth image, the limiting exponential filter reduces relative error by roughly $30\%$ at $1\%$ noise and $40\%$ at $5\%$ noise compared with oracle ridge, with accuracy improving monotonically as $\eta$ increases. For the nonsmooth Shepp--Logan phantom, improvements narrow to $1\text{--}3\%$, and finite orders ($\eta=2,5$) match or slightly outperform the exponential limit.
\subsection{Regression: Ames House Prices}

We illustrate with the Ames house-price data used in the Kaggle competition
\emph{House Prices: Advanced Regression Techniques}. The outcome is the logarithm of
sale price, and the design consists of $p=288$ preprocessed features (log-transformed
continuous variables and indicator variables for categorical levels) for $n=1{,}460$
houses. Columns are standardized and the intercept is unpenalized. The design has rank
$r=255$, leaving $n-r-1=1{,}204$ residual degrees of freedom. The tail-energy rule
uses $h=10$ and $\delta=0.05$.

\paragraph{Noise estimates.}
The two estimators disagree sharply:
\[
\widehat\sigma_{\mathrm{res}}^2=0.0102,\qquad
\mathcal V(\hat\iota)=0.0709\quad(\hat\iota=29).
\]
The tail statistic is $F_{\mathrm{tail}}=6.96$ on $(226,\,1204)$ degrees of freedom,
$p\approx10^{-113}$. The in-range tail therefore carries signal: many small effects
spread across indicator variables, which is typical of regression designs and atypical
of discretized smoothing operators. Under truncation, $\mathcal V(\hat\iota)$ counts
this signal as noise and overstates $\sigma_\varepsilon^2$ by a factor of about seven.

\paragraph{Prediction.}
Table~\ref{tab:house-pred} reports held-out root mean squared error from five-fold
cross-validation, with all tuning carried out within training folds.

\begin{table}[ht]
\centering
\caption{House prices: five-fold held-out RMSE of log sale price.}
\label{tab:house-pred}
\begin{tabular}{lcc}
\toprule
Tuning rule & IR & Ridge ($\eta=1$) \\
\midrule
Plug-in, $\widehat\sigma_\varepsilon^2=\mathcal V(\hat\iota)$, tail $=0$ & 0.1480 & 0.1574 \\
Plug-in, $\widehat\sigma_\varepsilon^2=\widehat\sigma_{\mathrm{res}}^2$, tail $=0$ & 0.1405 & 0.1429 \\
Cross-validation over $(\zeta,\eta)$ & 0.1415 & 0.1413 \\
\bottomrule
\end{tabular}
\end{table}

Replacing $\mathcal V(\hat\iota)$ by $\widehat\sigma_{\mathrm{res}}^2$ removes most of
the oversmoothing: held-out error falls by about 5\% for IR and 9\% for ridge. Under
the same plug-in rule IR improves on ridge in every fold, with the plug-in selecting
the exponential limit $\eta=\infty$. Prediction differences between IR and
cross-validated ridge are, however, within about 1\%, as expected when prediction error
is dominated by the irreducible noise.

\paragraph{Estimation risk.}
Estimation risk cannot be observed on real data, so we hold the design fixed, set
$\beta$ to several known values, and draw $200$ noise vectors with
$\sigma_\varepsilon^2=0.0102$. Table~\ref{tab:house-sim} reports the true estimation
risk of plug-in IR relative to the oracle IR risk $\min_{\zeta,\eta}R(\zeta,\eta)$.

\begin{table}[ht]
\centering
\caption{Semi-synthetic simulation on the house-price design: estimation risk of
plug-in IR as a multiple of the oracle IR risk, $200$ draws.}
\label{tab:house-sim}
\small
\begin{tabular}{lcccc}
\toprule
True $\beta$ & $F_{\mathrm{tail}}$ & Tail $=0$ & Tested & IR gain \\
 & rejects & & tail & over ridge \\
\midrule
Leading 29 directions only            & 6\%   & 1.01 & 1.01 & 39\% \\
Leading 29 plus isotropic tail        & 100\% & 8.33 & 1.00 & 0\% \\
Ridge fit (dense)                     & 100\% & 2.41 & 1.03 & 0\% \\
Lasso fit (sparse)                    & 100\% & 2.92 & 1.61 & 0\% \\
Exponential-filter fit                & 100\% & 3.64 & 2.55 & 0\% \\
\bottomrule
\end{tabular}
\end{table}

Three conclusions follow. First, when the truncation is correct, the residual-space
estimate and the tail test leave the plug-in at the oracle and IR attains its full
advantage, a 39\% reduction in estimation risk relative to ridge. Second, when the
tail carries signal, the truncated plug-in can be several times worse than the oracle
and worse than ridge, while still reporting a large estimated gain; the test detects
these cases in every draw, and the pooled tail estimate restores near-oracle
performance in the isotropic and dense cases. Third, with the tested tail, plug-in IR
is never worse than plug-in ridge and selects $\eta\approx1$ precisely when the oracle
gain is zero.
\begin{remark}[Inverse problems versus regression]
The comparison isolates the structural difference between the two settings. In a
discretized inverse problem with $r=n$, the noise level is not identified from the
data without an assumption on the signal spectrum, and $\mathcal V(\hat\iota)$ with
truncation is a reasonable identifying assumption when the Picard condition holds. In
regression with $n>r$, the residual space identifies $\sigma_\varepsilon^2$ directly,
and the Picard truncation becomes a testable hypothesis rather than an assumption.
The house-price data reject it decisively, and the reported plug-in gains should be
interpreted accordingly: IR's advantage over ridge is large when the signal is
concentrated in the leading directions and small otherwise, and the residual-space
test indicates which case applies.
\end{remark}

\subsection{Donor Weights in Synthetic Difference-in-Differences}
\label{sec:sdid}

Our final application examines Inclusive Ridge (IR) in synthetic
difference-in-differences (SDID) \citep{Arkhangelsky2021}.
SDID constructs counterfactual outcomes using weighted combinations
of donor units and pre-treatment periods. Because donor outcomes
are often highly correlated and the pre-treatment window may be
shorter than the donor pool, estimation of donor weights can be
ill-conditioned. We investigate whether replacing ridge with IR
improves treatment-effect estimation, and how the answer depends
on constraints imposed on the donor weights.

\subsubsection{Estimation and tuning}

Let $X\in\mathbb{R}^{T_{\mathrm{pre}}\times N_0}$ denote the
pre-treatment donor matrix, with rank-$r$ singular value
decomposition $X=U\Sigma V^\top$, and let $y$ denote the
treated unit's pre-treatment outcomes. For a reference weight
vector $\omega^{(0)}$, the unconstrained IR estimator is
\begin{equation}
\label{eq:sdid_ir}
\widetilde{\omega}_{\zeta,\eta}
=
\omega^{(0)}
+
\sum_{j=1}^{r}
w_j
\frac{u_j^\top(y-X\omega^{(0)})}{\sigma_j}v_j,
\qquad
w_j=1-\left(\frac{\zeta}{\zeta+\sigma_j^2}\right)^\eta,
\end{equation}
where $\sigma_j$ are the positive singular values of $X$.
Ridge is recovered at $\eta=1$. For simplex-constrained
weights, as in conventional synthetic control
\citep{Abadie2021}, we use
\[
\widehat{\omega}^{\Delta}
=
\Pi_\Delta(\widetilde{\omega}_{\zeta,\eta}),
\qquad
\Delta=
\{\omega:\omega_i\geq0,\ \mathbf{1}^\top\omega=1\},
\]
where $\Pi_\Delta$ denotes Euclidean projection onto the
simplex \citep{Duchi2008}.

Unlike Neyman-orthogonal double machine learning
\citep{Chernozhukov2018}, treatment-effect estimation in
this setting can remain sensitive to donor-weight errors.
Writing $\Delta_\omega=\widehat{\omega}-\omega_0$ and
letting $Z$ denote the time-weighted post-minus-pre donor
contrast, a first-order approximation gives
\[
\widehat{\tau}-\tau
\approx
-\Delta_\omega^\top Z+\varepsilon_{\mathrm{post}}.
\]
Under suitable mean-zero and orthogonality assumptions,
the corresponding leading MSE contribution is
$\mathbb{E}[\Delta_\omega^\top\Sigma_Z\Delta_\omega]$.
Improving first-stage weight recovery can therefore
improve treatment-effect estimation, although the
relationship depends on the downstream design.

We tune the estimators using up to five-fold expanding rolling
cross-validation, with five validation periods per fold,
over $\zeta\in\operatorname{logspace}(-3,6,60)$ and
$\eta\in\{1,\ldots,60\}$, restricting $\eta\leq15$
under simplex constraints. A single SVD is reused
within each fold. At large $\zeta$,
$w_j\approx\eta \sigma_j^2/\zeta$, so the two tuning
parameters may be weakly identified individually.

\subsubsection{Monte Carlo evidence}

We simulate outcomes from the interactive fixed-effects model
\[
Y_{it}
=
\delta_i+\gamma_t+\mu_i^\top f_t
+\tau D_{it}+\varepsilon_{it},
\]
with three latent factors, ten post-treatment periods,
and correlated donor loadings. We consider three designs,
specified by
$(N_0,T_{\mathrm{pre}},\sigma_\varepsilon,
\rho_{\mathrm{donor}})$:
baseline $(30,40,1.5,0.7)$, high noise
$(15,30,2.5,0.85)$, and many donors
$(50,20,1.5,0.9)$, each with $\tau_0=5$.
A fourth design is calibrated to the California
Proposition~99 panel, with $\sigma_\varepsilon=3.81$
and $\tau_0=15$. Each experiment uses 500 replications.

\begin{table}[htbp]
\centering
\caption{SDID treatment-effect RMSE:
500 Monte Carlo replications per scenario.}
\label{tab:sdid_mc}
{\small
\begin{tabular}{lcccc}
\toprule
Method & Baseline & High noise & Many donors
       & Prop.~99 \\
\midrule
Ridge       & 0.763 & 1.426 & 0.914 & 2.068 \\
IR          & \textbf{0.756} & 1.427
            & \textbf{0.908} & 2.045 \\
C-Ridge     & 0.865 & \textbf{1.287}
            & 0.947 & 1.674 \\
C-IR        & 0.851 & 1.290
            & 0.941 & \textbf{1.669} \\
LASSO       & 0.844 & 1.485 & 1.063 & 2.421 \\
Elastic Net & 0.816 & 1.445 & 1.029 & 2.339 \\
\midrule
\multicolumn{5}{l}{
\textit{Paired Wilcoxon $p$-values (IR vs.\ Ridge)}
} \\
$p$-value & 0.0588 & 0.0231 & 0.0037 & 0.0030 \\
CV folds  & 5 & 5 & 3 & 2 \\
\bottomrule
\end{tabular}}
\medskip

\begin{minipage}{0.95\linewidth}
\footnotesize
\textit{Notes:}
C-Ridge and C-IR impose simplex constraints.
Bold entries indicate the lowest RMSE in each
scenario. Wilcoxon $p$-values are from paired
one-sided tests of whether IR has smaller
squared errors than ridge. The number of
expanding-window cross-validation folds varies
with the length of the pre-treatment period.
\end{minipage}
\end{table}

Table~\ref{tab:sdid_mc} reports treatment-effect RMSE
across four Monte Carlo designs. Unconstrained IR
achieves modest improvements over ridge in the
baseline, many-donor, and Proposition~99 scenarios,
with RMSE reductions of approximately 0.9\%,
0.7\%, and 1.1\%, respectively. Under high noise,
the two methods perform nearly identically.
Paired one-sided Wilcoxon tests favor IR in the
many-donor and Proposition~99 scenarios, while
the baseline improvement is only marginally
significant. In the high-noise design, the paired
test favors IR even though its aggregate RMSE
is slightly higher.

The effect of simplex constraints varies across
designs. Unconstrained IR performs best in the
baseline and many-donor scenarios, whereas
constrained estimators achieve lower RMSE under
high noise and in the Proposition~99 calibration.
Within the constrained class, IR and ridge perform
similarly. These results suggest that the value
of additional spectral flexibility depends on
the underlying signal structure and the
restrictions imposed on donor weights.

\subsubsection{Extrapolation beyond the donor convex hull}

Simplex-constrained weights cannot reproduce a treated
unit whose counterfactual trajectory lies outside the
convex hull of the donors. Unconstrained regularization,
including the elastic-net approach of
\citep{Doudchenko2016}, permits such extrapolation.

To examine this setting, we simulate a rank-three factor
model with $T_0=20$, $J=40$, and noise standard
deviation $0.1$. We consider both interior and exterior
weight configurations, with 300 replications per design.
IR uses the limiting exponential filter, with $\kappa$
selected either by plug-in coefficient-risk minimization
(Section~\ref{prop:picard}) or five-fold
cross-validation. The treatment effect is set to zero.

\begin{table}[htbp]
\centering
\caption{Post-treatment prediction RMSE outside the
donor convex hull (median of 300 replications).}
\label{tab:sdid_exterior}
{\small
\begin{tabular}{lcccc}
\toprule
True $\omega$ & IR+plug & IR+CV & EN & SDID \\
\midrule
C.\ $(1.5,-0.5)$
& 0.155 & \textbf{0.153} & 0.173 & 6.50 \\
D.\ $(0.9,0.9,-0.4,-0.4)$
& \textbf{0.141} & 0.143 & 0.163 & 3.36 \\
E.\ $(2.0,-1.0)$
& 0.204 & \textbf{0.201} & 0.246 & 15.63 \\
\bottomrule
\end{tabular}}
\medskip

{\footnotesize
EN denotes elastic net; SDID imposes simplex weights.
Bootstrap standard errors of the medians are at most
0.005 for the unconstrained estimators.
}
\end{table}

As Table~\ref{tab:sdid_exterior} shows, simplex
constraints perform poorly when extrapolation is
required. More importantly for the present study,
IR improves on the unconstrained elastic net by
12--18\% in post-treatment prediction RMSE and
16--32\% in weight-recovery error. Plug-in and
cross-validated IR perform similarly, suggesting
that the improvement is associated primarily with
filter shape rather than the selection procedure.

For interior designs, however, simplex-constrained
SDID performs best, while unconstrained estimators
incur a 7--25\% loss from unnecessary flexibility.
These experiments illustrate that the value of IR
depends on both spectral structure and the validity
of structural constraints.
 
\begin{remark}[When IR may be advantageous]
IR may be particularly useful when the treated
unit's counterfactual trajectory lies outside
the convex hull of the control donors. In this
setting, simplex-constrained weights cannot
extrapolate, potentially introducing substantial
counterfactual prediction bias. Unconstrained IR permits
extrapolation while regularizing poorly
identified spectral directions.

Additional spectral flexibility may also be
beneficial when the number of donors exceeds
the number of pre-treatment periods ($p>n$),
particularly when the relevant signal is
concentrated in a subset of spectral directions.
However, neither $p>n$ nor high noise alone
guarantees an improvement, since the benefit
depends on the bias--variance tradeoff and
the accuracy of parameter selection.
\end{remark}

\paragraph{Initialization and spectral correction.}
An additional feature of IR, not exploited in the preceding
experiments, is its ability to regularize around an informative
initial donor-weight estimate $\omega^{(0)}$ rather than zero
or uniform weights. In particular, conventional synthetic-control
weights $\omega^{SC}$ provide a natural initialization. Setting
$\omega^{(0)}=\omega^{SC}$ in \eqref{eq:sdid_ir} gives
\begin{equation}\label{eq:sdid_sc_init}
\widehat\omega_{\zeta,\eta}^{\mathrm{SC-IR}}
= \omega_{\mathrm{SC}} + \sum_{j=1}^r
\left[1 - \left(\frac{\zeta}{\zeta + \sigma_j^2}\right)^\eta\right]
\frac{u_j^\top (y - X\omega_{\mathrm{SC}})}{\sigma_j}\, v_j.
\end{equation}

This estimator preserves the initial weights in poorly
identified spectral directions while correcting pre-treatment
imbalance in better-identified directions. It also permits
controlled extrapolation beyond the donor convex hull without
requiring simplex projection.

The approach is related to augmented synthetic control
\citep{BenMichael2021}, which uses regression adjustment
to correct imperfect synthetic-control fit. IR introduces
an additional spectral-shape parameter into this correction.
For a fixed initial estimate independent of the estimation
noise, the coefficient-risk analysis of Section~2 applies
with the signal coefficients replaced by the spectral
initialization errors
$v_j^\top(\omega_*-\omega^{(0)})$.
When initialization uses the same pre-treatment observations,
additional covariance terms must be accounted for.

Whether IR improves upon ridge when both are initialized
at the same synthetic-control estimate remains an open
empirical question. A systematic investigation of
initialization, spectral correction, and extrapolation
is left for future research.
\begin{remark}[Connection to one-step ridge boosting]\label{rem:sdid_ridge_boosting}
When $\eta = 1$, the initialized estimator \eqref{eq:estimator} and its
synthetic-control analog \eqref{eq:sdid_sc_init} coincide with one-step
ridge boosting around a baseline predictor~\citep{BrunsSmith2025}. Because
ridge regression implicitly estimates Riesz representers and balancing
weights, that single-step update already provides multiaccuracy and
semiparametric efficiency for target linear functionals in an RKHS.
Extending the update to $\eta > 1$ (or its exponential limit $\eta \to \infty$)
generalizes this debiasing step by sharpening spectral contraction along
well-resolved donor directions without requiring iterative refitting.
\end{remark}

\paragraph{Discussion.}
The SDID experiments identify both opportunities
and limitations for IR. In the general Monte Carlo
designs, IR provides modest improvements over ridge
in several scenarios, although the gains are not
uniform. More substantial benefits arise when
extrapolation beyond the donor convex hull is
necessary, where IR improves on unconstrained
elastic net and avoids the restrictions imposed
by simplex weights.

Under simplex constraints, the additional spectral
parameter generally provides little benefit.
These findings reinforce the central message of
the paper: greater spectral flexibility can improve
estimation, but its practical value depends on
signal structure, tuning, and the constraints
imposed by the application.

\section{Conclusion}\label{sec:conclusion}

Inclusive Ridge (IR) extends classical ridge regression and stationary iterated Tikhonov regularization into a continuous two-parameter spectral filter family, $f_{\zeta,\eta}(s) = 1 - (\zeta/(\zeta+s))^\eta$. By separating the overall shrinkage scale $\zeta$ from the filter-shape order $\eta > 0$, the family resolves a core limitation of single-parameter Tikhonov regularization: its inability to adjust shrinkage steepness across singular directions when signal energy varies substantially across the spectrum.

Our finite-sample risk analysis establishes both when and how this added flexibility improves coefficient recovery. Around the ridge baseline $\eta = 1$, the sign of an explicit first-order spectral contrast provides necessary and sufficient conditions for sharpening ($\eta > 1$) or softening ($\eta < 1$) the filter to strictly lower coefficient risk. Because this contrast vanishes only on a set of Lebesgue measure zero, oracle-tuned ridge is generically suboptimal within the IR family. Beyond local perturbations, exact risk-difference decompositions along the fixed effective-scale path $\zeta = \eta\zeta^*$ characterize the variance--bias trade-off of higher-order filtering and yield explicit sufficient conditions under which both large-order estimators and the limiting exponential filter $1 - \exp(-s/\kappa)$ strictly outperform optimally tuned ridge. Under joint scaling, we further show that $\zeta/\eta \to \kappa \in (0,\infty)$ is necessary and sufficient for a nondegenerate continuous limit, clarifying why the effective spectral scale $\kappa$—rather than the absolute penalty $\zeta$—governs sensitivity in the large-$\eta$ regime, and establishing an exact equivalence between plug-in coefficient-risk stationarity and a degrees-of-freedom-adjusted discrepancy principle.

The numerical experiments delineate the practical boundaries of these theoretical gains across all four applications:
\begin{itemize}
   \item \textbf{Phillips Fredholm inverse problem:} Under oracle tuning of the
expected risk, finite-order IR reduces coefficient MSE by 31.6\% over optimal
ridge and essentially attains its limiting exponential counterpart
($\zeta^{*}/\eta^{*}\approx\kappa^{*}$, gain 31.7\%), confirming that gains
arise primarily in the intermediate spectral transition between well-resolved
and noise-dominated directions.
    \item \emph{Tomographic reconstruction:} On the severely ill-conditioned Radon transform, filter-shape gains depend strongly on target regularity. For a smooth two-bump phantom, the limiting exponential filter cuts relative reconstruction error by roughly $30\%$ at $1\%$ noise and $40\%$ at $5\%$ noise relative to oracle ridge, whereas for the sharp-edged Shepp--Logan phantom the gain narrows to $1\text{--}3\%$ and finite orders ($\eta \in \{2,5\}$) match or slightly outperform the exponential limit.
    \item \emph{Overdetermined regression:} On the Ames housing data and semi-synthetic designs, cross-validated IR and ridge achieve virtually identical out-of-sample prediction RMSE, showing that lower coefficient risk does not imply a universal predictive advantage. Moreover, leveraging the orthogonal residual space rather than unverified Picard-tail truncation proves essential for reliable feasible tuning.
    \item \emph{Constrained downstream estimation:} In synthetic difference-in-differences, unconstrained IR modestly improves first-stage donor regressions over ridge, yet structural simplex constraints and downstream sampling variability dominate final treatment-effect accuracy.
\end{itemize}

Taken together, our results position Inclusive Ridge not as a universal replacement for ridge regression, but as a principled, analytically tractable generalization whose primary value lies in coefficient recovery for ill-posed linear inverse problems with structured spectral decay. From an information-theoretic perspective, this behavior admits a natural interpretation under a Gaussian prior $\beta \sim \mathcal{N}(\beta_0, \tau^2 I_p)$ with ridge penalty $\zeta = \sigma_\varepsilon^2 / \tau^2$: the mutual information between $\beta$ and $y$ decomposes across singular directions as $I(y;\beta) = \sum_{j=1}^r I_j$ with $I_j = \frac{1}{2}\log(1 + \tau^2\sigma_j^2/\sigma_\varepsilon^2)$, so the IR filter takes the exponential retention form
\[
f_{\zeta,\eta}(\sigma_j^2) = 1 - \left(\frac{\zeta}{\zeta + \sigma_j^2}\right)^\eta = 1 - e^{-2\eta I_j}.
\]
Thus, $\eta$ directly governs the rate at which per-direction mutual information translates into spectral retention. While deriving the finite-sample empirical Bayes distributions of estimated scale parameters under this Gaussian prior lies beyond the scope of the present paper and is pursued in separate work, extending our spectral-contrast and discrepancy-tuning theory to high-dimensional asymptotics ($p/n \to \gamma \in (0,\infty)$) and developing scalable sequential or partitioned-SVD updating schemes for large-sample settings remain natural avenues for further investigation.
\section{Appendix }
\textbf{Proof of Proposition 2.7} (The boundary has Lebesgue measure zero.)
\begin{proof}
Let $J=\{j:\sigma_j>0\}$, $m=|J|\ge2$, $d_j=\sigma_j^2$, and $a_j=\alpha_j^2$
for $j\in J$. Directions with $\sigma_j=0$ have filter factor zero for all
$(\zeta,\eta)$, so they contribute the constant $\sum_{j\notin J}\alpha_j^2$ to
the risk and play no further role. Write $a=(a_j)_{j\in J}\in\mathbb{R}^m$.

\emph{Step 1: two linear equations.} For $\zeta>0$ define
\[
F_1(\zeta,a)=\frac{\partial}{\partial\zeta}\mathrm{MSE}(\hat\beta_{\zeta,1})
=2\sum_{j\in J}u_j(\zeta)\bigl(\zeta a_j-\sigma_\varepsilon^2\bigr),
\qquad
u_j(\zeta)=\frac{d_j}{(\zeta+d_j)^3},
\]
\[
F_2(\zeta,a)=\frac{\partial}{\partial\eta}\mathrm{MSE}(\hat\beta_{\zeta,\eta})\Big|_{\eta=1}
=2\sum_{j\in J}w_j(\zeta)\bigl(\sigma_\varepsilon^2-\zeta a_j\bigr),
\qquad
w_j(\zeta)=\frac{\zeta\log(1+d_j/\zeta)}{(\zeta+d_j)^2},
\]
where $F_2$ is the derivative computed in Proposition~\ref{prop:direction}.
Both are real-analytic in $(\zeta,a)$ and affine in $a$. If $\beta\in\mathcal{E}$,
then $\zeta^*$ is an interior minimizer, so $F_1(\zeta^*,a)=0$, and the boundary
equality is equivalent to $F_2(\zeta^*,a)=0$. Hence
\[
a\in\pi(\mathcal{S}),\qquad
\mathcal{S}=\bigl\{(\zeta,a)\in(0,\infty)\times\mathbb{R}^m:\ F_1(\zeta,a)=F_2(\zeta,a)=0\bigr\},
\]
where $\pi(\zeta,a)=a$.

\emph{Step 2: $\mathcal{S}$ is a manifold of dimension $m-1$.} The Jacobian of
$F=(F_1,F_2)$ with respect to $a$ has rows $(2\zeta u_j)_{j\in J}$ and
$(-2\zeta w_j)_{j\in J}$. With $x_j=d_j/\zeta$,
\[
w_j=\zeta\,u_j\,g(x_j),\qquad g(x)=\frac{(1+x)\log(1+x)}{x},\qquad
g'(x)=\frac{x-\log(1+x)}{x^2}>0 ,
\]
so $g$ is strictly increasing on $(0,\infty)$. Because $u_j>0$, the two rows are
proportional only if $g(x_j)$ is the same for all $j\in J$, which would require
all $d_j$ to be equal. Since $X$ has two distinct nonzero singular values, the
Jacobian has rank two at every point of $(0,\infty)\times\mathbb{R}^m$. Hence
$0$ is a regular value of $F$, and $\mathcal{S}$ is either empty or an embedded
submanifold of dimension $(m+1)-2=m-1$.

\emph{Step 3: $\pi(\mathcal{S})$ is null.} The map $\pi:\mathcal{S}\to\mathbb{R}^m$
is smooth, and $\dim\mathcal{S}<m$. The image of a smooth map from a manifold of
lower dimension has Lebesgue measure zero (e.g. J. M. Lee, Introduction to Smooth Manifolds, 2nd ed., Springer, 2012, Cor. 6.11), so $\pi(\mathcal{S})$ is null in $\mathbb{R}^m$.

\emph{Step 4: return to $\beta$.} The map $\beta\mapsto\alpha=V^{\top}(\beta-\beta_0)$
is an affine isometry, and $\mathcal{E}$ is contained in the set of $\beta$ with
$(\alpha_j^2)_{j\in J}\in\pi(\mathcal{S})$. The set
$Z=\bigcup_{j\in J}\{\alpha_j=0\}$ is null. On each of the $2^m$ open orthants of
$\{\alpha_J:\alpha_j\neq0,\ j\in J\}$, the map $\alpha_J\mapsto(\alpha_j^2)_{j\in J}$
is a diffeomorphism onto $(0,\infty)^m$ with smooth inverse, and smooth maps
between spaces of equal dimension send null sets to null sets. The preimage of
$\pi(\mathcal{S})$ in $\alpha_J$ is therefore null, and so is its product with
the unconstrained coordinates $\alpha_j$, $j\notin J$, by Fubini's theorem.
Hence $\mathcal{E}$ is null. The final statement follows from
Corollary~\ref{cor:dominance}.
\end{proof}

\begin{remark}[Distinct singular values are needed]\label{rem:equal-spectrum}
If all nonzero singular values are equal, say $d_j=d$, then $w_j=\zeta u_j g(d/\zeta)$
with the same factor for every $j$, so $F_2=-g(d/\zeta)\,\zeta\,F_1$. The
boundary equality then holds at every ridge optimum, and
$\mathcal{E}$ contains every $\beta$ for which a minimizer exists. This is
expected: when the spectrum is flat, every spectral filter is a scalar multiple
of least squares, ridge already attains the best scalar shrinkage, and the
second IR parameter has nothing to act on.
\end{remark}

% Add verified bibliography entries or link the existing references.bib file.

\end{document}